\documentclass{ws-procs9x6}

\newcommand{\beq} {\begin{equation}}
\newcommand{\eeq} {\end{equation}}
\newcommand{\beqa} {\begin{eqnarray}}
\newcommand{\eeqa} {\end{eqnarray}}
\newcommand{\mrm}[1] {{\mathrm{#1}}}

\newcommand{\ie}{{\it i.e.}}
\newcommand{\eg}{{\it e.g.}}

\newcommand{\gev}{{\mrm{\ GeV}}}

\newcommand{\order}[1]{${\mathcal O}\left(#1 \right)$}
\newcommand{\morder}[1]{{\mathcal O}\left(#1 \right)}
\newcommand{\eq}[1]{(\ref{#1})}

\newcommand{\halft}{{\textstyle \frac{1}{2}}}

\newcommand{\gsim}{\gtrsim}

\begin{document}

\title{Summary Talk on Quark-Hadron Duality\footnote{\uppercase{W}ork partially
supported by the \uppercase{A}cademy of \uppercase{F}inland through grant 102046.}}

\author{Paul Hoyer}

\address{Department of Physical Sciences and Helsinki Institute of Physics \\
POB 64, FIN - 00014 University of Helsinki, Finland\\
and\\
NORDITA, Blegdamsvej 17, DK-2100 Copenhagen, Denmark\\
E-mail: paul.hoyer@helsinki.fi}

\maketitle

\abstracts{
I ascribe the origin of Bloom-Gilman duality in DIS to a separation of scales between the hard subprocess and soft resonance
formation. The success of duality indicates that the
subprocesses of exclusive form factors are the same as in DIS.
The observed dominance of the longitudinal structure function at large $x$ in
$\pi N \to \mu^+\mu^- X$ can explain why local duality works for DIS with a pion target.
The failure of duality in semi-exclusive processes indicates that high momentum transfer $t$ is not sufficient to make the corresponding subprocesses compact.}

\vspace{-12cm}
\phantom{0}    
\phantom{0}    

\hfill HIP-2005-37/TH

\hfill NORDITA-2005-55

\vspace{10cm}

This meeting\footnote{First Workshop on Quark-Hadron Duality and the Transition to pQCD, Laboratori Nazionali di Frascati, June 6-8 2005; http://www.lnf.infn.it/conference/duality05/ .} demonstrated the exciting progress made in the last few years on duality in Deep Inelastic Scattering (DIS). Building on the work of Bloom and Gilman\cite{Bloom:1970xb} 35 years ago, high precision data principally from Jlab\cite{Niculescu:2000tj,Niculescu:2000tk} and DESY\cite{Airapetian:2002rw} has reopened the field, allowing detailed studies of duality including spin dependence and nuclear effects. In this written summary I focus on just a few aspects of duality that I think give clues to the underlying QCD dynamics. I refer to the presentations given at the workshop for the many important results that I cannot cover here. The comprehensive review\cite{Melnitchouk:2005zr} by Melnitchouk, Ent and Keppel covers the experimental and theoretical results on duality available before the workshop.

\section{Duality and the uncertainty relation} \label{uncertainty}

The duality between resonances and hard perturbative processes is most easily visualized in $e^+e^-$ annihilations, where vector mesons average the asymptotic $e^+e^- \to Q\bar Q$ cross section\footnote{This is, in particular, exploited in the ``QCD Sum Rules''\cite{Shifman:1978bx}.}. Quarks with large mass $M_Q$ are produced at a short time-scale $1/M_Q$. Resonances form at a longer time-scale $1/\Delta M_{Q\bar Q}$ characterized by their mass differences. Thus resonance formation is {\it incoherent with the hard subprocess}, \ie, it cannot affect the quark production probability. At the time of resonance formation the total energy uncertainty $\Delta E \sim \Delta M_{Q\bar Q}$ limits the range within which the perturbative cross section may be ``reshuffled'' into resonance peaks.

The local duality between single resonances and the perturbative $e^+e^-$ cross section can thus be viewed as a consequence of the relative softness of the interactions involved in resonance formation. In more differential quantities, such as spin-dependent structure functions, the scale $\Delta M$ refers to mass differences between resonances having the same spin dependence. Thus semilocal duality in DIS is found to work separately for the $N(940)$ and $\Delta(1232)$ in spin-averaged cross sections, whereas a broader averaging region appears to be required for spin-dependent quantities\cite{Melnitchouk:2005zr,ent,blok}. The larger $\Delta M$ may also explain why BG duality sets in at a higher value of $Q^2$ in the spin-dependent structure function, since the subprocess and resonance scales need to be clearly separated. 

\begin{figure}[ht]
\centerline{\epsfxsize=8cm\epsfbox{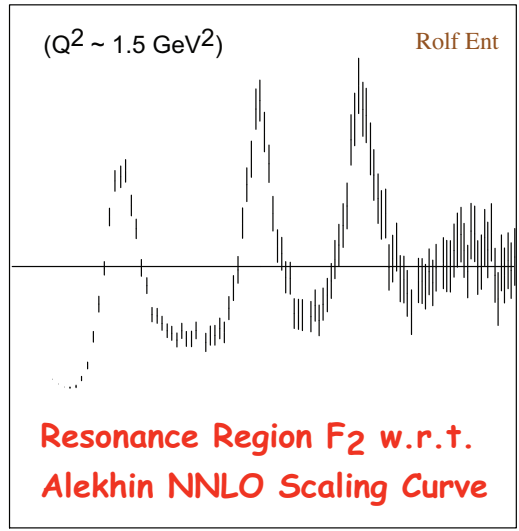}}   
\caption{Comparison\protect\cite{ent} between the $x$-dependence of the $F_2(x,Q^2)$ structure function measured at low $Q^2 \sim$ 1.5 GeV$^2$ (errror bars with resonance structures) and the scaling curve measured at high $Q^2$ (horizontal line). \label{fig1}}
\end{figure}

\section{Duality in Deep Inelastic Scattering}

Compared to $e^+e^-$, BG duality in DIS opens up a new dimension: At each value of $Q^2$ there is a whole range of Bjorken $x$-values in which the asymptotic cross section may be compared to the resonance contributions (Fig.~1). As $Q^2$ increases, a given resonance $N^*$ of mass $M$ contributes at an increasing value of $x = Q^2/(Q^2+M^2)$, thus ``sliding'' along the scaling curve $F_2(x)$ towards $x=1$. If the transition form factor $F_{N\to N^*}(Q^2) \sim 1/Q^{2n}$ then local duality requires that the inclusive quark distribution behaves as $F_{q/N}(x) \sim (1-x)^{2n-1}$ for $x\to 1$.

\begin{figure}[ht]
\centerline{\epsfxsize=4.1in\epsfbox{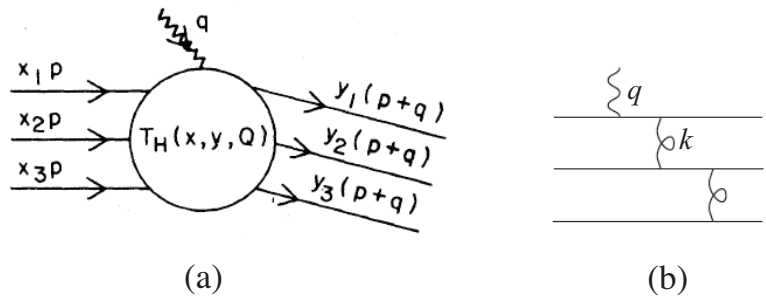}}   
\caption{QCD form factor dynamics according to Lepage and Brodsky\protect\cite{bl:1980}. A generic diagram of the hard subprocess $T_H$ in (a) is shown in (b). The momenta $k$ of the gluon exchanges scale with the photon momentum $q$. \label{fig2}}
\end{figure}

In DIS the virtual photon scatters incoherently from each quark in the target. Hence the cross section depends on the sum of squared quark charges,
\beq \label{insum}
F_2 \sim \sum_q e_q^2
\eeq
Exclusive form factors (Fig.~2) are assumed\cite{bl:1980} be built from target Fock states whose transverse dimensions are compatible with the photon resolution. The gluons exchanged in the hard subprocess shown in Fig.~2(b) then have momenta $k \sim \morder{Q}$, and the photon couples coherently to all quarks,
\beq \label{exsum}
F_{N\to N^*} \sim \left(\sum_q e_q \right)^2
\eeq
The electric charges of the quarks are unrelated to QCD dynamics. The different weighting in \eq{insum} and \eq{exsum} thus appears incompatible with BG duality. The observation\cite{Close:2001ha} that the interference terms in \eq{exsum} cancel when averaging over resonances with different parity is not sufficient, since duality works within each resonance region (Fig.~1).

In the previous Section we saw that duality can be understood as a consequence of the unequal time scales involved in the hard subprocess and resonance formation. Once the hard process has ``happened'', later interactions can only redistribute the (inclusive) cross section within a limited mass range. Such an explanation requires that the hard subprocesses in resonance form factors and in the scaling DIS cross section {\it are the same}. In particular, the Fock states contributing to form factors must have a transverse size exceeding the photon resolution, so that coherent scattering from several quarks is suppressed.

PQCD calculations of proton and pion form factors\cite{Bolz:1994hb} indicate that the virtuality $k^2$ of the gluons in Fig.~2(b) is typically much smaller than the photon virtuality $Q^2$. This means that the photon will effectively couple incoherently to the quarks, and the weighting of quark charges will be according to \eq{insum}, just as in DIS. 

The fact that BG duality works furthermore indicates that the hard subprocess has reached its scaling limit at the moderate $Q^2$ value of Fig.~1. If the resonances build the scaling distribution they must dominate the inclusive cross section at the corresponding value of $x$ (given by the resonance mass). The $\Delta(1232)$ actually decreases faster with $Q^2$ than the scaling cross section -- but the difference is taken up by ``background''\cite{Carlson:1993wy}. Thus BG duality still works locally in the $\Delta(1232)$ region, by relating the scaling (high $Q^2$) curve to the combined production of resonance and background at lower $Q^2$. This again requires that the subprocess has reached its scaling limit (in $Q^2$ at fixed $x$), whereas the $\Delta(1232)$/background ratio, which is determined by the soft hadronization process, decreases as $x \to 1$.

\section{The longitudinal structure function}

BG duality has been found to work\cite{Melnitchouk:2005zr,ent} also in the longitudinal photon structure function $F_L(x,Q^2)$ for $Q^2 \gsim 1.5 \gev^2$. This is an important check, as subprocesses involving longitudinal photons differ from those of transverse photons (\eg, longitudinal photons do not couple to on-shell spin $\halft$ quarks at leading twist). The success of duality again indicates that resonances are produced by the same hard subprocesses as those responsible for the scaling longitudinal structure function.

\begin{figure}[htb]
\centerline{\epsfxsize=4.1in\epsfbox{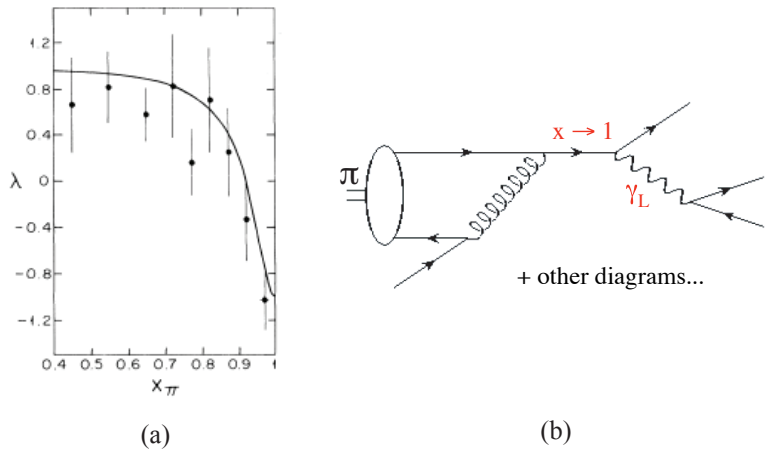}}   
\caption{(a) The parameter $\lambda$ of the angular distribution of the muons in $\pi N \to \mu^+ \mu^- X$, as a function of the fractional momentum $x_\pi$ of the projectile pion carried by the muon pair\protect\cite{Palestini:1985zc}. $\lambda = +1\ (-1)$ corresponds to a transversely (longitudinally) polarized virtual photon. The solid line is from Ref.\protect\cite{Berger:1979du}. (b) A generic diagram contributing in the limit $Q^2 \to \infty$ with $Q^2(1-x)=M^2$ held fixed. The $x \to 1$ quark propagator has virtuality of \order{Q^2} and is thus coherent with the virtual photon.
 \label{fig3}}
\end{figure}

The pion elastic form factor (measured by $e\pi \to e\pi$) gets a contribution only from longitudinal photon exchange. One would thus expect that the pion be dual to $F_L$ measured on a pion target. While no DIS data on pion targets exist, the pion structure function has been measured in the Drell-Yan process $\pi N \to \mu^+ \mu^- X$. The muon angular distribution shows\cite{Palestini:1985zc} that the longitudinal structure function dominates at large $x$ (Fig.~3a). This may be understood as a consequence of helicity conservation -- the photon carries the helicity of the projectile as its momentum fraction $x_\pi \to 1$. Longitudinal photons can dominate over transverse photons in the limit $Q^2 \to \infty,\ x \to 1$ with $M^2 = Q^2(1-x)$ fixed since diagrams like the one in Fig.~3(b) contribute at leading order \cite{Berger:1979du,Brodsky:1991dj}. This limit is appropriate for the contribution of a given resonance of mass $M$ to DIS. It is, however, quite different from the standard Bjorken limit where $x$ is held fixed, the twist expansion is relevant and transverse photons dominate.

\begin{figure}[htb]
\centerline{\epsfxsize=9cm\epsfbox{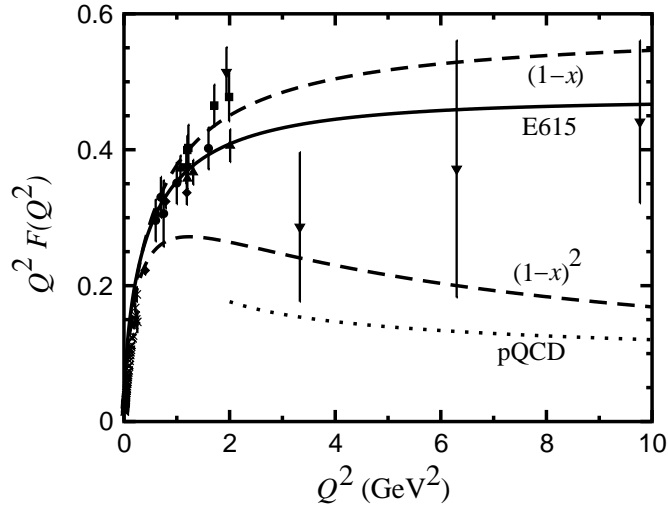}} 
\caption{Duality comparison\protect\cite{Melnitchouk:2002gh} of the elastic pion form factor (data points) with the E615 quark distribution of the pion measured in the Drell-Yan process. \label{fig4}}
\end{figure}

The effective quark distribution in the pion was found\cite{Palestini:1985zc} to behave as $f_{q/\pi}(x) \sim (1-x)^{1.12\pm 0.18}$ at high $x$. In QCD one expects\cite{Farrar:1975yb} $f_{q/\pi}(x) \sim (1-x)^2$ for transverse photons in the Bjorken limit. The data is not in conflict with QCD due to the contribution of longitudinal photons. The pion elastic form factor is expected\cite{bl:1980} to behave as $F_\pi(Q^2) \sim 1/Q^2$ in QCD, which is consistent with the available data. Local BG duality for the pion then requires $f_{q/\pi}(x) \sim (1-x)^1$, tantalizingly close to the data. In fact, even the normalization is consistent with local duality for the pion\cite{Melnitchouk:2002gh}, as shown in Fig.~4. This agreement lends further support to the dominance of the longitudinal structure function in the data at high $x$.

\section{Extending duality: Semi-exclusive processes}

I have argued that BG duality provides a tool for understanding the dominant dynamics of exclusive form factors. It is obviously important to try to generalize duality beyond DIS. Promising applications to semi-inclusive processes were already discussed at this workshop\cite{ent,bosted}. Here I shall mention a somewhat different approach which turns out not to work -- but the failure is striking enough to be instructive.

\begin{figure}[htb]
\centerline{\epsfxsize=4.1in\epsfbox{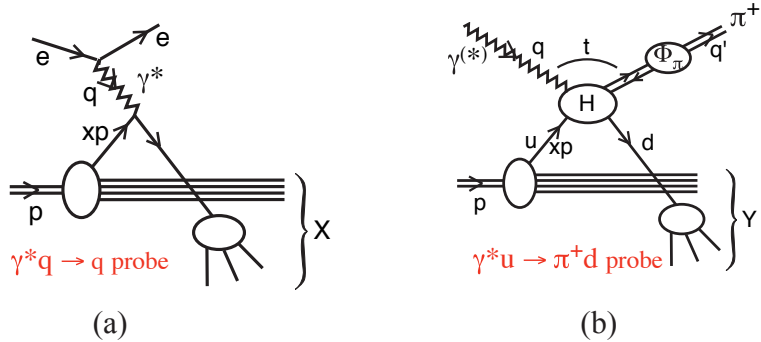}}   
\caption{Similarity of dynamics in DIS (a) and semi-exclusive pion production (b). The hard subprocess $eq\to eq$ of DIS is replaced with the compact $\gamma^{(*)}u \to \pi^+ d$ subprocess in the semi-exclusive process. \label{fig5}}
\end{figure}

As indicated in Fig.~5(a), in DIS we use the hard (PQCD calculable) subprocess $eq \to eq$ to probe the quark structure of a nucleon target. BG duality than implies that we can relate the inclusive cross sections measured at the same value of $x$ but different $Q^2$ (and hence also different masses of the inclusive system $X$, down to the resonance region). Analogously, in Fig.~5(b) we are probing the quark distribution using a different subprocess $\gamma^{(*)}u \to \pi^+ d$. Insofar as this subprocess is hard, the quark pair forming the $\pi^+$ will be produced in a compact, color singlet configuration and will not further interact in the target due to color transparency. In a kinematical limit where the hadrons in the inclusive system $Y$ are separated by a large rapidity gap from the $\pi^+$ we may calculate the cross section of this {\it semi-exclusive process} using PQCD and the standard parton distributions\cite{Carlson:1993ys}. The soft dynamics forming the inclusive system $Y$ is again incoherent with the hard subprocess and we may relate cross sections at various $M_Y$ by appealing to duality. 

An application of the above idea to $\gamma p \to \pi^+ n$ at large momentum transfer $t$ gave an interestingly incorrect result\cite{Eden:2001ci}. There is no data on the semi-inclusive process $\gamma p \to \pi^+ Y$, but we expected to get a (ball-park) estimate for the exclusive process with $Y=n$ using local duality analogously to DIS. However, the calculation turned out to be off by nearly two orders of magnitude! The corresponding analysis of wide-angle Compton scattering $\gamma p \to \gamma p$ did not fare much better, underestimating the data by about a factor 10. To me the most likely explanation is that the subprocess $H$ in Fig.~5(b) actually is soft even at high $t$ (when the incoming photon is real). Clearly we have much to learn about the dynamics of processes involving large momentum transfers -- and Bloom-Gilman duality provides us with a very powerful tool.

\section*{Acknowledgments}

I am very grateful to the organizers for arranging this most topical workshop, and for asking me to give a contribution.

\end{document}